\documentclass[reprint,amsmath,twocolumn,amssymb, nobibnotes, aps, pra, superscriptaddress]{revtex4-1}

\usepackage{graphicx}
\usepackage{dcolumn}
\usepackage{bm}
\usepackage{color}
\usepackage{tikz}
\usetikzlibrary{shapes}
\usetikzlibrary{backgrounds}
\usetikzlibrary{plotmarks}

\begin{document}
\def \beq{\begin{equation}}
\def \eeq{\end{equation}}
\def \bea{\begin{eqnarray}}
\def \eea{\end{eqnarray}}
\def \bem{\begin{displaymath}}
\def \eem{\end{displaymath}}
\def \P{\Psi}
\def \Pd{|\Psi(\boldsymbol{r})|}
\def \Pds{|\Psi^{\ast}(\boldsymbol{r})|}
\def \Po{\overline{\Psi}}
\def \bs{\boldsymbol}
\def \bl{\bar{\boldsymbol{l}}}
\newcommand{\ihat}{\hat{\textbf{\i}}}
\newcommand{\jhat}{\hat{\textbf{\j}}}

\title{The Haldane Model in a Magneto-optical Honeycomb  Lattice}

\author{Mark J. Ablowitz}
\affiliation{Department of Applied Mathematics, University of Colorado, Boulder, Colorado, USA}
\author{Justin T. Cole}
\affiliation{Department of Mathematics, University of Colorado, Colorado Springs, Colorado, USA}
\date{\today}     

\begin{abstract}

A two-dimensional honeycomb lattice composed of gyrotropic rods is studied.  Beginning with Maxwell’s equations, a perturbed Wannier method is introduced which yields a tight-binding model with nearest and next-nearest neighbors. The resulting discrete model  leads to a Haldane model and as such, topologically protected modes, associated with nonzero Chern numbers are supported.  Changing the radii of the rods allows for the breaking of inversion symmetry which can change the topology of the system. This model explains experimental results associated with topological waves in magneto-optical honeycomb lattices. This method  can also be applied to more general Chern insulator lattices. When on-site Kerr type nonlinear effects are considered, coherent soliton-like modes are found to propagate robustly through boundary defects.


\end{abstract}
\maketitle

\section{Introduction} 

The study of topological insulators is an area of research currently receiving significant interest. These types of systems can be experimentally realized in numerous fields including ultracold fermionic systems \cite{Jotzu2014}, semiconductors \cite{vonKlitzing1980,Bernevig2006}, magnetic media \cite{Chang2013}, equatorial waves \cite{Delplace2017}, and electromagnetic systems \cite{Rechtsman2013,Ozawa2019,Lu2014}. Underlying  these works are topologically protected states that are robust to defects.

This work focuses on topological insulators that are distinguished by bulk eigenmodes with a nontrivial Chern number.
In this case, the bulk-edge correspondence implies the existence of topologically protected modes. 
Indeed, these systems can support edge states that propagate unidirectionally around the boundary with and without material defects.

A standard approach for describing topological insulator lattice systems is the use of a tight-binding model. Typically, tight-binding models consist of a set of discrete equations that reduce the complexity of the governing equations, yet still capture the essential behavior. Moreover, it is common in experiments for the dielectric contrast in photonic waveguides to naturally reside in the deep lattice regime which is central in the tight-binding approximation \cite{Ablowitz2022a,Fefferman2018}.

One of the most well-known and heavily studied topological insulator systems is the {\it Haldane model} \cite{Haldane1988}, associated with honeycomb lattices. This relatively simple model, which includes nearest
and next-nearest neighbor interactions, is able to capture the essence of Chern insulator systems. 
The model illustrates that breaking of time-reversal symmetry is necessary, but not sufficient, for realizing bulk modes with nontrivial Chern topological invariants. Moreover, when inversion symmetry is broken in an appropriate manner exceeding that of time-reversal symmetry, a topological transition to a trivial Chern system can take place. 

While  \cite{Haldane1988} offers no derivation for the 
model, it effectively describes the behavior of the quantum Hall effect in  honeycomb lattices.  Indeed, many authors have applied the Haldane model to describe systems with nonzero Chern numbers. However, it is not clear whether this is the true tight-binding reduction of electromagnetic systems or just a convenient model.

This work provides  a direct derivation of the Haldane model from Maxwell's equations in a magneto-optical (MO) system. The physical system considered here is that of transverse magnetic (TM) waves in a  ferrimagnetic photonic crystal with an applied external magnetic field.  
Systems of this type have been realized in both square \cite{Wang2008,Wang2009,Sun2019,Yang2013} and honeycomb \cite{Ao2009,Poo2011,Zhao2020} lattices and found to support topologically protected edge modes. Topologically protected modes in EM systems were originally proposed by Haldane and Raghu \cite{Haldane2008,Raghu2008} and their existence studied in \cite{LeeThorp2019}.  Our work directly connects Haldane's model \cite{Haldane1988} to electro-magneto-optical systems. 
Additionally, it turns out that periodically driven photonic honeycomb lattices can
also yield the Haldane model  in a
high-frequency limit \cite{Jotzu2014,Ablowitz2022}. 

The key to our approach is to use a suitable Wannier  basis 
in which to expand the EM field \cite{Ablowitz2020}. Unfortunately, a  direct Wannier expansion  is ineffective due to  nontrivial topology which is the result of a discontinuity  in the spectral phase of the associated Bloch function \cite{Brouder2007,Ablowitz2022}. As a result, the corresponding Wannier-Fourier coefficients do not decay rapidly. Seeking a tight-binding model  in a basis of these slowly decaying Wannier modes would require many interactions, well beyond nearest neighbor, to accurately describe the problem. Consequently, this would cease to be an effective reduction of the original problem. 

By considering nearest and next-nearest neighbor interactions, a Haldane model is derived from the original MO honeycomb lattice. 
With  physically relevant parameters, an analytical study of the system topology is conducted. The topological transition points are identified and found to agree well with numerical approximations. Nontrivial Chern numbers are found to correspond to unidirectional chiral modes, and vice versa for trivial Chern cases.

 The Wannier basis method we use was applied to a square MO  lattice in \cite{Ablowitz2020}. The results in this paper show that the method  is effective, again. This approach can be applied to other systems, e.g. different lattices, governed by the TM equation with gyrotropic lattices. We expect this method to model other Chern insulator systems, e.g. TE systems with gyrotropic permittivity, as well.

We also examine the effect of nonlinearity on edge mode propagation. Edge solitons, unidirectional nonlinear envelopes that balance nonlinearity and dispersion have been  explored in Floquet Chern insulator systems \cite{Ablowitz2014,Ablowitz2017}. The work \cite{Mukherjee2021}  showed that significant amounts of radiation are emitted from the solitary wave for highly localized (nonlinear) envelopes. 

The nonlinear system we consider is a Haldane model 
that includes on-site Kerr nonlinearity. A similar system has also been derived from nonlinear a Floquet system in a high-frequency driving limit \cite{Ablowitz2022}. Different    nonlinear Haldane models with saturable nonlinearity  \cite{Harari2018} and  mass terms \cite{Zhou2017} have previously been explored. Using
balanced envelopes such as those described above as a guide, we observe that slowly-varying envelopes can propagate coherently and robustly around lattice boundaries. Due to their ability to balance nonlinear and dispersive effects, while localized along the boundary, 
we call these  edge solitons.

\section{Magneto-optical System} 

\begin{figure}
\centering
\includegraphics[scale=.6]{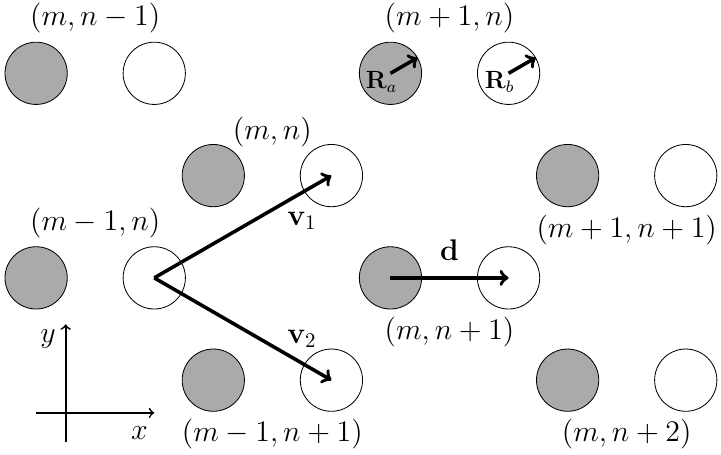}
 \caption{Planar honeycomb array of ferrite rods. The lattice a-sites (b-sites) are  grey (white) discs with radius $R_a$ ($R_b$). A constant magnetic field $H_0 \widehat{{\bf z}}$ is applied. 
 The indices $(m,n)$ denote integer shifts in the ${\bf v}_1$ and ${\bf v}_2$ directions, respectively. The displacement between  two rods in the same  cell is ${\bf d}$ with $||{\bf d}||_2 = \ell$. \label{HC_lattice}}
\end{figure}

The setup we consider is a planar array ${\bf r} = (x,y)^T$ of ferrimagnetic rods, e.g. YIG rods,  arranged in a honeycomb lattice pattern (see Fig.~\ref{HC_lattice}). Similar designs were implemented in \cite{Wang2008,Wang2009} and \cite{Ao2009,Poo2011}.  The parallelogram unit cell contains an ``a-site'' and ``b-site'' with radii of $R_a$  and $R_b$, respectively. All other cells are integer translations of the lattice vectors 
$${\bf v}_1 = \ell  \begin{pmatrix} \frac{3}{2} \\ \frac{\sqrt{3}}{2} \end{pmatrix} , ~~~ {\bf v}_2 = \ell  \begin{pmatrix} \frac{3}{2} \\ -\frac{\sqrt{3}}{2} \end{pmatrix} $$
from the unit cell, where $\ell$ is the distance between nearest neighbor rods. The notation $(m,n)$ indicates a rod that is displaced $m {\bf v}_1 + n {\bf v}_2$ away from the unit cell, where $m,n \in\mathbb{Z}.$

A constant external magnetic field is applied in the perpendicular (out of the page) direction: $H_0 \widehat{\bf z}$, and saturated magnetization. 
For time-harmonic fields with angular frequency $\omega$, the ferrite rods induce the gyrotropic permeability tensor \cite{Pozar} 
\begin{equation*}
[\mu] = 
\begin{pmatrix}
\mu & i \kappa & 0 \\
- i \kappa & \mu & 0 \\
0 & 0 & \mu_0
\end{pmatrix} ,
\end{equation*}
where $\mu = \mu_0 \left( 1 + \frac{\omega_0 \omega_m}{ \omega_0^2 - \omega^2} \right)$ and $\kappa = \mu_0 \frac{\omega \omega_m}{\omega_0^2 - \omega^2}$. The coefficients are defined in terms of $\omega_0 = \mu_0 \gamma H_0$ and $\omega_m = \mu_0 \gamma M_s$, where $\mu_0$ is the vacuum permeability, $\gamma$ is the gyromagnetic ratio, and $M_s$ is the magnetization saturation of the material. 

For rods with  permittivity $\varepsilon({\bf r})$, the governing TM wave equation for a time-harmonic field is
\begin{eqnarray}
\label{TM_eqn}
& -\nabla^2 E + \mathcal{M} \cdot \nabla E = \omega^2 \varepsilon \tilde{\mu} E , & \\
\nonumber
& \mathcal{M}({\bf r}) =  \nabla \ln \tilde{\mu} - i \tilde{\mu} (\widehat{{\bf z}} \times \nabla \eta) , &
\end{eqnarray}
where $E$, plus its complex conjugate, is the $z$-component of the electric field, $\tilde{\mu} = \frac{ \mu^2 - \kappa^2}{\mu}$ and $\eta = - \frac{\kappa}{\mu^2 - \kappa^2}$. Here we  take a non-dispersive approximation and fix the values of $\mu$ and $\kappa$: $\omega$ is fixed to eventual band gap frequencies. For a typical YIG rod at frequency $f = 7.7$GHz ($f=\omega/2\pi$)  with saturation magnetization $4 \pi M_s = 1750 $ G and magnetizing field $H_0 = 500 $ Oe, the constitutive relations are approximately $\mu = 0.88 \mu_0, \kappa = -0.66 \mu_0 $ and $\varepsilon = 15 \varepsilon_0$. The equation is nondimensionalized via: ${\bf r} \rightarrow \ell {\bf r}, \mu \rightarrow \mu_0 \mu, \kappa \rightarrow \mu_0 \kappa, \varepsilon \rightarrow \varepsilon_0 \varepsilon, \omega \rightarrow c\omega/\ell$, where $c$ is the speed of light and $\varepsilon_0 = (c^2 \mu_0)^{-1}$ is the vacuum permittivity.

The coefficients in (\ref{TM_eqn}) share the translation symmetry of the honeycomb lattice: $\varepsilon({\bf r} + m {\bf v}_1 + n {\bf v}_2) = \varepsilon({\bf r} ), \tilde{\mu}({\bf r} + m {\bf v}_1 + n {\bf v}_2) = \tilde{\mu}({\bf r} ),$ and $\eta({\bf r} + m {\bf v}_1 + n {\bf v}_2) = \eta({\bf r} )$ where $m,n \in \mathbb{Z}$. Bloch theory motivates bulk wave solutions of 
the  form $E({\bf r} ; {\bf k})  = e^{ i {\bf k} \cdot {\bf r}}u({\bf r}; {\bf k})$, where $u({\bf r}+ m {\bf v}_1 + n {\bf v}_2 ; {\bf k})= u({\bf r}; {\bf k})$ for quasimomentum ${\bf k}$ where these 
reciprocal lattice vectors are given by
$${\bf k}_1 = \frac{2 \pi}{\ell} \begin{pmatrix} \frac{1}{3} \\\ \frac{1}{\sqrt{3}} \end{pmatrix}, ~~~ {\bf k}_2 = \frac{2 \pi}{\ell} \begin{pmatrix} ~ \frac{1}{3} \\\ - \frac{1}{\sqrt{3}} \end{pmatrix}  .$$

\begin{figure}
\includegraphics[scale=.375]{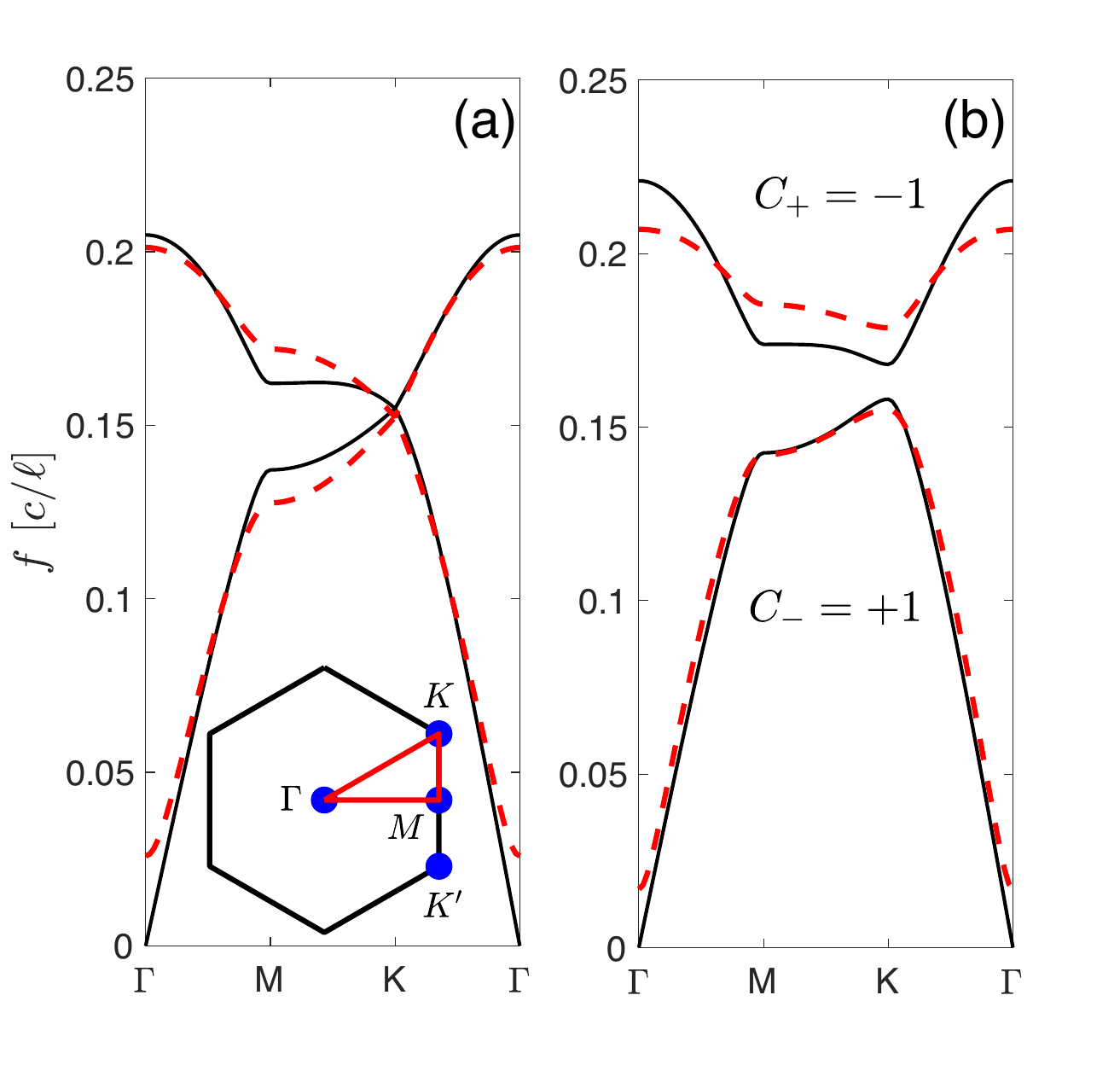}
 \caption{Normalized frequency bands obtained by  solving  (\ref{TM_eqn}) for $u({\bf r}; {\bf k}) = e^{- i {\bf k} \cdot {\bf r}} E({\bf r}; {\bf k}) $
 along the path $\Gamma M K \Gamma$ with $R_a = R_b = 0.3 \ell$. Shown are the first two bands (ascending order) for (a) non-magnetized ($ \mathcal{M} = {\bf 0}, \tilde{\mu} = \mu_0$) and (b) magnetized ($ \mathcal{M} \not= {\bf 0}, \tilde{\mu} \not= \mu_0$) systems. Solid lines indicate numerically computed curves. 
 Dashed lines denote the tight-binding approximation discussed below. \label{bulk_bands_compare}}
\end{figure}

Solving  the resulting equation for $(u({\bf r}; {\bf k}), \omega({\bf k}))$,
via spectral methods along the $\Gamma M K \Gamma$ path in ${\bf k}-$space (see Appendix~\ref{numerical_compute_bands}), we obtain the  two lowest spectral bands shown in Fig.~\ref{bulk_bands_compare}.
The location of Dirac points are 
$$K' = \begin{pmatrix} 0 \\  \frac{4 \pi}{3 \sqrt{3}\ell} \end{pmatrix} , ~~ K = \begin{pmatrix} 0 \\ - \frac{4 \pi}{3 \sqrt{3}\ell} \end{pmatrix}.$$

In Fig.~\ref{bulk_bands_compare}(a), no external magnetic field is applied and a conical Dirac point is observed at the $K$ point. When a magnetic field is applied, then $\mathcal{M}$ is nonzero, time-reversal symmetry is broken and a band gap opens [see Fig.~\ref{bulk_bands_compare}(b)]. Moreover, there is an associated set of nonzero Chern numbers. Note that the first (lowest) band is denoted by `--' subscript, while the second band is denoted by `+' subscript. 
In  Fig.~\ref{bulk_bands_compare} we also compare with the discrete--tight binding approximation discussed below.

\section{A Perturbed Wannier Approach} 
\label{perturbed_wannier_approach}

 A strong dielectric contrast between the rods and background  motivates a tight-binding approximation, whereby a variable coefficient PDE with a periodic lattice potential, i.e. (\ref{TM_eqn}), can be reduced to a constant coefficient system of ODEs  \cite{Ablowitz2022a}. Bloch wave solutions of (\ref{TM_eqn}) are periodic with respect to the quasimomentum ${\bf k}$:  $ E({\bf r}; {\bf k} + m {\bf k}_1 + n {\bf k}_2 )  = E({\bf r}; {\bf k})$,
where the reciprocal lattice vectors ${\bf k}_{1,2}$ and satisfy ${\bf v}_i \cdot {\bf k}_j = 2 \pi \delta_{ij}$. As such, the Bloch wave can be expanded in a Fourier in $ {\bf k}$
series
\begin{equation}
\label{Bloch_wave}
E({\bf r}; {\bf k}) = \sum_{p} \sum_{m,n}  W_{mn}^p({\bf r}) e^{i {\bf k} \cdot (m {\bf v}_1 + n {\bf v}_2) } , 
\end{equation}
where $W_{mn}^p$ denotes the Wannier function corresponding to the $(m,n)$ spatial cell and $p^{\rm th}$ spectral band. 

Due to the properties of Fourier coefficients, the decay of $W_{mn}^p({\bf r})$ depends on the smoothness of $E({\bf r}; {\bf k})$ in ${\bf k}$. Chern insulators possess an essential phase discontinuity that can not be removed via gauge transformation \cite{Brouder2007}. As a result,  a direct Wannier expansion is not useful. But, a closely related set of exponentially localized Wannier functions, which come from a problem {\it with} time-reversal symmetry \cite{Ablowitz2020} can be used perturbatively.

\begin{figure}
\includegraphics[scale=.4]{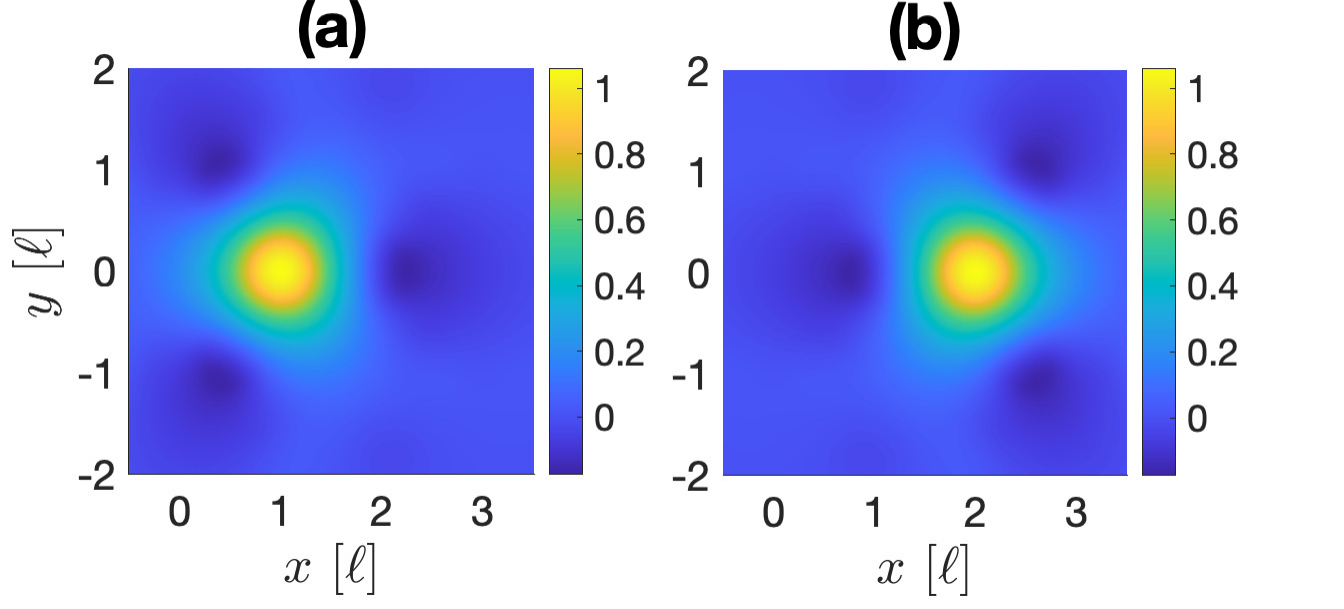}
 \caption{Maximally localized Wannier functions (a) $\widetilde{W}^a_{00}({\bf r})$ and (b)$\widetilde{W}^b_{00}({\bf r})$, obtained from (\ref{TM_eqn}) 
 when $\mathcal{M} = {\bf 0}$ and $R_a = R_b = 0.3 \ell$. \label{Wannier_Fcn_Plot}}
\end{figure}

Consider (\ref{TM_eqn}) with $ \mathcal{M} = {\bf 0}$, but $\tilde{\mu}\not=\mu_0$; the so-called ``perturbed problem''.
The maximally localized Wannier (MLWF) functions, found using well-known methods  \cite{Marzari1997}, corresponding to the first two Wannier functions,
called $\widetilde{W}^{a}_{mn}({\bf r})$ and $\widetilde{W}^{b}_{mn}({\bf r})$, are shown in Fig.~\ref{Wannier_Fcn_Plot} and centered at the a-sites (${\bf d} + m {\bf v}_1 + n {\bf v}_2$) and b-sites ($2{\bf d} + m {\bf v}_1 + n {\bf v}_2$), respectively. These Wannier functions are related to those in (\ref{Bloch_wave}) by a unitary transformation chosen to  minimize the variance 
(see Appendix~\ref{appendix_break_inversion}). 
Note that these functions are  real, exponentially localized, and approximately possess
mirror symmetry about the $x = 3 \ell /2$ axis; i.e. $\widetilde{W}^a_{00}(- ({\bf r} - 3\ell \widehat{{\bf x}}/2 )) = \widetilde{W}^b_{00}( {\bf r} - 3\ell \widehat{{\bf x}}/2 )$.

The Bloch wave is expanded in terms of this new basis as 
\begin{align}
\label{Bloch_expand}
{E}({\bf r}; {\bf k}) &=  \sum_{m,n} (a_{mn} \widetilde{W}^a_{mn}({\bf r}) + b_{mn} \widetilde{W}^b_{mn}({\bf r})) ,
\end{align}
where the phases have been absorbed into the coefficients. Properly normalized Wannier modes exhibit the orthogonality property $\langle W_{mn}^p , W_{m'n'}^{p'} \rangle_{\mathbb{R}^2} = \delta_{mm'} \delta_{n n'} \delta_{pp'} $ for the weighted complex inner product $\langle f , g \rangle_{\mathbb{R}^2} = \iint_{\mathbb{R}^2} f({\bf r})^* g({\bf r}) \varepsilon({\bf r}) \tilde{\mu}({\bf r}) d {\bf r}$.

Substituting (\ref{Bloch_expand}) into (\ref{TM_eqn}) with $\mathcal{M} \not= 0$, multiplying by $\widetilde{W}^j_{mn}({\bf r}), j = a,b,$ and integrating over $\mathbb{R}^2$ yields a system of algebraic equations whose coefficients depend on 
integrals over perturbed Wannier functions. Once the MLWFs are obtained, these integrals are  numerically approximated.
Due to the deep lattice, in the simplest tight-binding approximation only nearby interactions are kept since the others are  small. Below 
we keep terms up to the next-nearest neighboring sites.

\section{A Haldane-type Model} 

 Inspection of the numerically computed tight-binding coefficients (see Appendix \ref{numerical_compute_bands})  reveals an
effective discrete approximation that is essentially 
 the well-known Haldane model \cite{Haldane1988}. Namely, replacing $\omega$ by $id/dt$ 
we obtain 
\begin{align}
\label{Haldane1}
\frac{d^2 a_{mn}}{dt^2} & + P a_{mn} + t_1 (\delta_- b_{mn} ) \\ 
\nonumber & + t_2  e^{i \phi} ( \Delta_1  a_{mn} ) +  t_2 e^{-i \phi} ( \Delta_2  a_{mn} ) = 0 \\
\label{Haldane2}
\frac{d^2 b_{mn}}{dt^2} & + \widetilde{P} b_{mn} + t_1 (\delta_+ a_{mn} ) \\
\nonumber & + \tilde{t}_2 e^{-i {\phi}} ( \Delta_1 b_{mn} )+ \tilde{t}_2 e^{i {\phi}} ( \Delta_2  b_{mn}) = 0 
\end{align}
where $(\delta_{\pm} c_{mn}) \equiv c_{mn} + c_{m \pm 1, n } + c_{m, n\pm 1}$ are the nearest neighbor interactions and $(\Delta_1 c_{mn}) \equiv c_{m,n+1} + c_{m-1,n} + c_{m+1,n-1}$ and $(\Delta_2 c_{mn}) \equiv c_{m+1,n} + c_{m,n-1} + c_{m-1,n+1}$ are  next-nearest neighbor contributions; 
parameters $P, \widetilde{P}, t_1, t_2, \tilde{t}_2, \phi$ are real numbers; these values depend on the values of $\mu, \kappa, \varepsilon$ and sizes of the radii of the rods.
Notice that this system reduces to the `classical' Haldane model given in \cite{Haldane1988} when $\tilde{t}_2 = t_2$ and $\phi \rightarrow - \phi$. The equations can be put in a more standard form by looking for solutions of the form $a_{mn} \rightarrow  a_{mn}e^{i \omega t},$
similarly for $b_{mn},$ and then shifting the spectrum
$\omega^2 \rightarrow \omega^2 + (P + \widetilde{P})/2$. This yields an on-site inversion parameter 
 \begin{equation}
 M \equiv \frac{P - \widetilde{P}}{2}
 \label{Meq}
 \end{equation}
which is important in \cite{Haldane1988} and below. The result of this latter spectral shift is to effectively replace $P$ in (\ref{Haldane1}) by $M$ and $\widetilde{P}$ in (\ref{Haldane2}) by $-M$.

We find the `classical' Haldane model when inversion symmetry is {\it not} broken ($P = \widetilde{P}, t_2 = \tilde{t}_2 )$ and a modified version when inversion symmetry {\it is} broken ($P \neq \widetilde{P}, t_2 \not= \tilde{t}_2)$. 
When the a-site and b-site rods differ, the inversion symmetry of the lattice ${\bf r} \rightarrow - {\bf r}$ is broken and this leads to different interactions among the Wannier modes (see Sec.~\ref{broken_inversion_symmetry} and Appendix~\ref{appendix_break_inversion}).

The physical derivation that leads to the model (\ref{Haldane1})-(\ref{Haldane2}) should be pointed out. The external magnetic field induces the complex next-nearest coefficients in the system. 
Furthermore, this method appears applicable for the derivation of other tight-binding models in Chern insulator systems.

We compare the  bulk bands of the discrete model to those numerically computed from (\ref{TM_eqn}); see Fig.~\ref{bulk_bands_compare}. (All tight-binding parameters used  
can be found in Appendix~\ref{tight_binding_parameters}) Indeed, the discrete approximation shows good agreement with the numerical bands; the relative error throughout the Brillouin zone is 6.5\% or less. Moreover, for $\ell = 5.8$ mm spacing, the gap frequencies in Fig.~\ref{bulk_bands_compare}(b) lie in the vicinity of the 8 GHz microwave regime
found in \cite{Poo2011}. 
 
\subsection{Analytical Calculation of Bulk Modes} 
\label{analytical_topology_calc}
 
 Consider bulk plane wave  solutions of system (\ref{Haldane1})-(\ref{Haldane2}) of the form
\begin{align*} 
a_{mn}(t) &= \alpha({\bf k}) e^{i \left[ {\bf k} \cdot (m {\bf v}_1 + n {\bf v}_2)  - \omega({\bf k}) t \right]}  , \\
b_{mn}(t) &= \beta({\bf k}) e^{i \left[ {\bf k} \cdot (m {\bf v}_1 + n {\bf v}_2  )  - \omega({\bf k}) t \right]} ,
\end{align*}
where ${\bf k} \in \mathbb{R}$. Next, define the nearest neighbor and next-nearest neighbor vectors
${\bf a}_1 = {\bf 0}, {\bf a}_2 =  {\bf v}_1 , {\bf a}_3 ={\bf v}_2  $,
and ${\bf b}_1 = {\bf v}_1 , {\bf b}_2 = - {\bf v}_2 , {\bf b}_3 = {\bf v}_2 - {\bf v}_1,$ respectively.	
Then the bulk Haldane system can be expressed as
the following eigenvalue system
\begin{equation}  \label{BulkHaldane}
\begin{pmatrix}
M + H_0 + H_3 & H_1 - i H_2 \\
H_1 + i H_2 & - M + \tau (H_0 - H_3)
\end{pmatrix}
\begin{pmatrix}
\alpha \\ \beta
\end{pmatrix}
= \widetilde{\omega}^2
\begin{pmatrix}
\alpha \\ \beta
\end{pmatrix}
\end{equation}
where $\tau = \tilde{t}_2/t_2 > 0$ and $M = (P - \widetilde{P})/2$ with the terms
\begin{align*}
H_0({\bf k}) & = 2 t_2 \sum_{j = 1}^3 \cos \phi \cos ( {\bf k} \cdot {\bf b}_j ) \\
H_1({\bf k}) & =  t_1 \sum_{j = 1}^3  \cos(  {\bf k} \cdot {\bf a}_j ) \\
H_2({\bf k}) & =  t_1 \sum_{j = 1}^3  \sin ( {\bf k} \cdot {\bf a}_j ) \\
H_3({\bf k}) & = 2  t_2 \sum_{j = 1}^3 \sin \phi  \sin ( {\bf k} \cdot {\bf b}_j ) .
\end{align*}
Note that we have utilized  the frequency shift $\omega^2 = \widetilde{\omega}^2 + (P + \widetilde{P})/2$ to follow the convention used in \cite{Haldane1988}.
When $\tau = 1$ this is precisely Haldane's model \cite{Haldane1988} when $\phi \rightarrow - \phi$ and $\widetilde{\omega}^2 \rightarrow \overline{\omega}^2  + H_0$.
The dispersion surfaces of (\ref{BulkHaldane}) are given by
\begin{widetext}
\begin{equation} 
 \label{BulkDispersion}
  \widetilde{\omega}^2_{\pm}({\bf k})  =  \frac{H_0({\bf k}) (1 + \tau) + H_3({\bf k}) (1 - \tau)}{2}   
 \pm \sqrt{ H_1({\bf k})^2 + H_2({\bf k})^2 + \frac{1}{4} \left[2M+  H_0({\bf k})(1 - \tau) + H_3({\bf k}) (1 + \tau) \right]^2 } .
\end{equation}
\end{widetext}

\begin{figure}
\centering
\includegraphics[scale=.38]{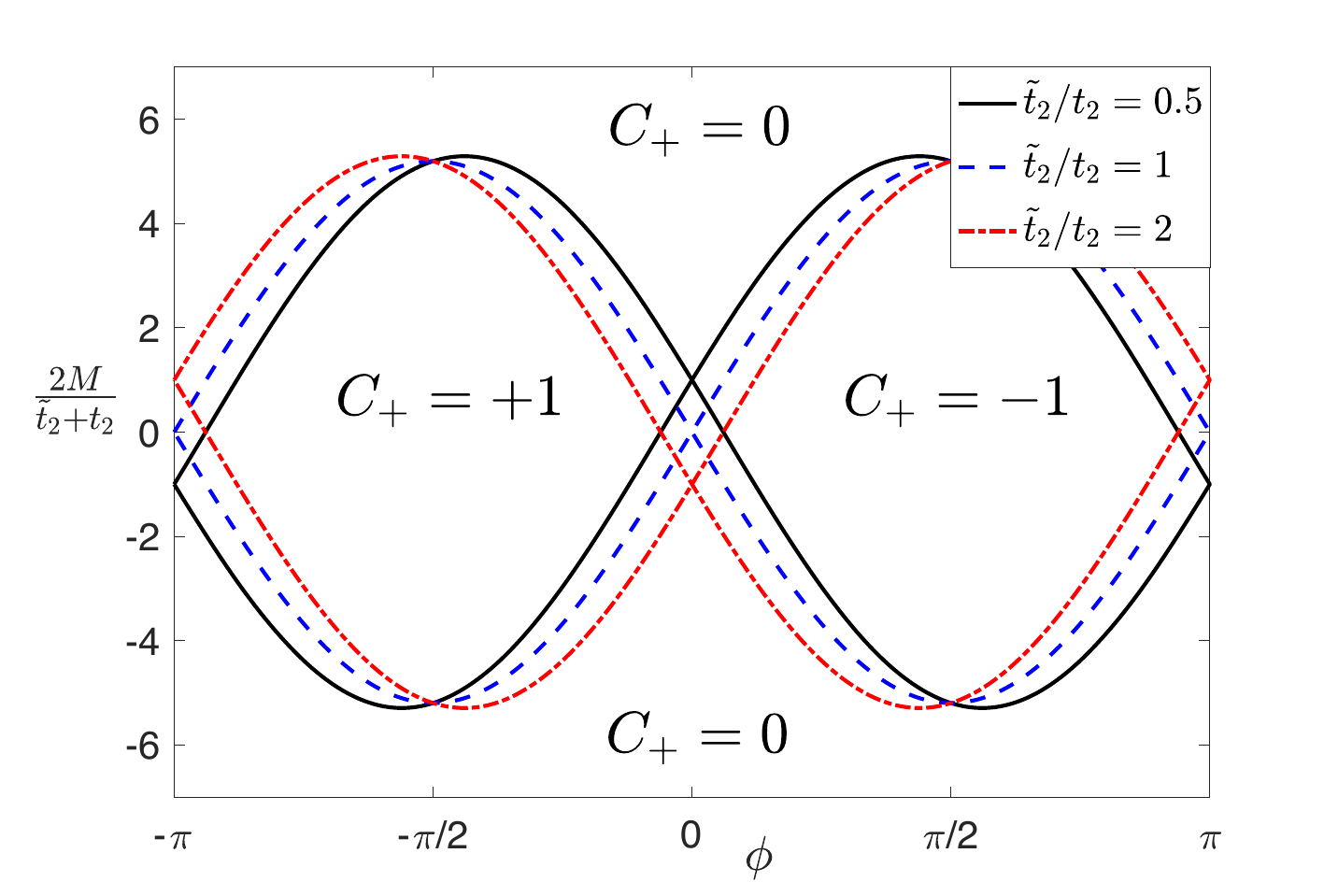}
 \caption{Phase diagram separating nontrivial Chern insulator (interior) and trivial  system (exterior) for different ratios of $\tilde{t}_2/ t_2.$ Shown is the Chern number $C_+$ 
 for the second spectral band; the signs are reversed for $C_-$ 
 corresponding to the first band. Note: $\tau = \tilde{t}_2/t_2$ is a function of 
 $R_a/R_b$. 
 \label{gap_touch_region}}
\end{figure}

Below, we begin by studying the behavior of the spectrum near the Dirac points. In the absence of magnetization ($t_2 = \tilde{t}_2 = 0$), the spectral gap closes and the bands $\widetilde{\omega}_{\pm}$ touch at these  points. Moreover, as will be explained below, the contributions that result in nonzero Chern numbers are acquired at these points.

Consider the behavior of the spectral bands in (\ref{BulkDispersion}) at the Dirac point $K' = ( 0 , \frac{4 \pi}{ 3\sqrt{3} \ell})^T$, where the functions $H_j, j=1,...,4$ reduce to 
\begin{align*}
& H_0(K')  =  - 3 t_2  \cos \phi, ~~   H_1(K')  =  0, \\
& H_2(K')  =  0, ~~  H_3(K')  =  3 \sqrt{3}  t_2 \sin \phi.
\end{align*}
Hence, at this Dirac point, the spectral bands in (\ref{BulkDispersion}) are given by 
\begin{align*}
 \widetilde{\omega}^2_{\pm} = & \frac{- 3 t_2 \cos \phi (1 + \tau) + 3 \sqrt{3} t_2 \sin \phi (1- \tau)}{2} \\
& \pm\frac{1}{2}\left| 2M - 3 t_2 \cos \phi (1 - \tau) +3 \sqrt{3} t_2 \sin \phi (1 + \tau)  \right| .
\end{align*}
A gap closure (i.e. $\widetilde{\omega}_{+}= \widetilde{\omega}_{-}$) occurs when the equation
\begin{align}
\label{band_touch_Kp}
2 M  - 3 (t_2 - \tilde{t}_2)  \cos \phi + 3 \sqrt{3} (t_2 + \tilde{t}_2)  \sin \phi = 0 
\end{align}
is satisfied.

If on the other hand, the Dirac point $K = - K'$ is considered, then
\begin{align*}
&H_0(K)  =  - 3 t_2  \cos \phi, ~~   H_1(K)  =  0, \\ 
&H_2(K)  =  0, ~~  H_3(K)  = - 3 \sqrt{3}  t_2 \sin \phi,
\end{align*}
where the only difference is the sign of $H_3$. Here, the corresponding gap closure occurs when the equation
\begin{align}
\label{band_touch_K}
 2 M  - 3 (t_2 - \tilde{t}_2)  \cos \phi - 3 \sqrt{3} (t_2 + \tilde{t}_2)  \sin \phi = 0
\end{align}
is satisfied. 
When  $\tau = 1$ (inversion symmetry present), the gap closure condition reduces to
that of the classical Haldane model: $M \pm 3 \sqrt{3}  \sin \phi = 0.$ The curves in (\ref{band_touch_Kp}) and (\ref{band_touch_K}) are shown in Fig.~\ref{gap_touch_region} for different values of $\tau$ and  correspond to topological transition points.

The  eigenmodes associated with the $\widetilde{\omega}_{\pm}^2$ eigenvalues  in (\ref{BulkDispersion}) are 
\begin{equation}
 \label{BulkEigMode}
{\bf c}_\pm({\bf k}) =   \frac{1}{D({\bf k})}
\begin{pmatrix}
 H_1({\bf k}) - i H_2({\bf k}) \\
 \widetilde{\omega}_\pm^2({\bf k}) -M - H_0({\bf k}) - H_3({\bf k})  
 \end{pmatrix} .
\end{equation}
The term $D({\bf k})$ is a normalization factor chosen to ensure $|| {\bf c}_{\pm} ||_2 =1.$
Notice that these functions are periodic in {\bf k}:	
${\bf c}_{\pm}({\bf k} + {\bf k}_j) = {\bf c}_{\pm}({\bf k} ) ,$
for the reciprocal lattice vectors, $j = 1,2$.

The Chern number for the first two bands is given by
\begin{equation}
 \label{ChernDefine}
C_{\pm} = \frac{1}{2 \pi i} \iint_{\Omega} \left( \bigg< \frac{\partial {\bf c}_\pm }{ \partial k_x} ,\frac{\partial {\bf c}_\pm }{ \partial k_y} \bigg> -  \bigg< \frac{\partial {\bf c}_\pm }{ \partial k_y} ,\frac{\partial {\bf c}_\pm }{ \partial k_x} \bigg>\right)  d {\bf k} ,
\end{equation}
where $\langle  {\bf f} , {\bf g} \rangle = {\bf f}^\dag {\bf g}$ and $\dag$ indicates the complex conjugate transpose. 
The region $\Omega$ is a reciprocal unit cell, given by the parallelogram region formed by the reciprocal lattice vectors ${\bf k}_1, {\bf k}_2$.

 To compute (\ref{ChernDefine}), Stokes' Theorem is applied over $\Omega$.
This equates the double integral  over $\Omega$ to a closed line integral along the boundary $\partial \Omega$.
However, since the eigenfunctions  (\ref{BulkEigMode}) are not differentiable at the Dirac points, a contour integral which excludes these  points must be implemented (see \cite{Ablowitz2022}).
Due to the periodic boundary conditions in the eigenmodes, the  boundary $\partial \Omega$ makes no contribution to the Chern number. The only nontrivial contributions  come from the two Dirac points:
\begin{equation}
 \label{ChernDiracPoints}
C_{\pm} = -  \frac{1}{2 \pi i} \oint_{\partial K} {\bf A}_{\pm}({\bf k}) \cdot d {\bf k} - \frac{1}{2 \pi i} \oint_{\partial K'} {\bf A}_{\pm}({\bf k}) \cdot d {\bf k} 
\end{equation}
where ${\bf A}_{\pm}({\bf k}) = \langle {\bf c}_\pm , \nabla_{\bf k} {\bf c}_\pm \rangle = \langle {\bf c}_\pm , \partial_{k_x} {\bf c}_\pm \rangle \widehat{k}_x + \langle {\bf c}_\pm , \partial_{k_y} {\bf c}_\pm \rangle \widehat{k}_y$ is the Berry connection.
The contours of integration  in (\ref{ChernDiracPoints}) are taken to be small counter-clockwise oriented  circles centered around the Dirac points, $K$ and $K'$, respectively.

Next, the eigenmodes are linearized about the Dirac point ${\bf k} = K'$. A similar calculation follows for the other Dirac point. Doing so, we get 
\begin{equation*}
{\bf c}_{\pm}({\bf k}) \approx {\bf c}_{\pm}(K') + ({\bf k} - K') \cdot \nabla_{\bf k} {\bf c}_{\pm}(K') ,
\end{equation*}
where $\nabla_{\bf k}  \equiv \partial_{k_x} \widehat{k}_x + \partial_{k_y} \widehat{k}_y$. 
After renormalizing the linear approximation via ${\bf \psi} = {\bf c}_{\pm}/|| {\bf c}_{\pm} ||_2$, the Berry connection and Chern number (\ref{ChernDefine}) are computed in the neighborhood of the $K'$ Dirac point.

The following are the results. The contribution to the total Chern number at the $K'$ Dirac point is $-1$ for 
\begin{equation}
 \label{ChernKp}
2 M  - 3 (t_2 - \tilde{t}_2)  \cos \phi + 3 \sqrt{3} (t_2 + \tilde{t}_2)  \sin \phi > 0
 \end{equation}
and $0$ otherwise. Meanwhile, the contribution  at the $K$ Dirac point is $+1$ for 
\begin{equation}
 \label{ChernK}
 2 M  - 3 (t_2 - \tilde{t}_2)  \cos \phi - 3 \sqrt{3} (t_2 + \tilde{t}_2)  \sin \phi > 0
 \end{equation}
and $0$ otherwise.  These different region of topology are summarized in Fig.~\ref{gap_touch_region}. This figure represents a generalization of the phase diagram in \cite{Haldane1988}.

The Chern number is found by combining the contributions  in (\ref{ChernKp}) and (\ref{ChernK}). Suppose we focus on the interval $0 < \phi < \pi$. Then for parameters that satisfy neither  (\ref{ChernKp}) nor (\ref{ChernK}),  that is $2 M  < 3 (t_2 - \tilde{t}_2)  \cos \phi - 3 \sqrt{3} (t_2 + \tilde{t}_2)  \sin \phi,$
both Dirac points have zero contribution and $C_+ =0 + 0 = 0.$ Next, for values  that satisfy (\ref{ChernKp}), but not (\ref{ChernK}), 
the $K'$ Dirac point contributes $-1$, and the $K$ point has a null contribution, so $C_+ = - 1 + 0 = -1$. Lastly, when both (\ref{ChernKp}) and (\ref{ChernK}) are satisfied, that is
$2 M  > 3 (t_2 - \tilde{t}_2)  \cos \phi + 3 \sqrt{3} (t_2 + \tilde{t}_2)  \sin \phi,$
both Dirac points contribute and cancel each other out, so $C_+ = -1 + 1 = 0.$ For all cases considered in this paper, the analytically computed Chern numbers agree with numerics \cite{fukui}.

Lastly,
it is observed that, similar to the classic Haldane model, bands only open for $\phi \not= n \pi, n \in \mathbb{Z}$. Physically,  values of $\phi = n \pi$ corresponds to completely real next-nearest neighbor coefficients. Opening a spectral gap that supports topologically protected edge modes
requires complex next-nearest neighbor coefficients. In this model, the complex nature of the next-nearest neighbor coefficients comes from the external magnetic field.


\subsection{Broken Inversion Symmetry} 
\label{broken_inversion_symmetry}

A notable feature of the Haldane model is a change in topology when the degree to which the inversion symmetry is broken is sufficiently large.
The generalized model (\ref{Haldane1})-(\ref{Haldane2}) also exhibits this property. Physically, inversion symmetry of the system can be broken by choosing different radii for the $a$ and $b$ lattice sites, that is $R_a \not= R_b$. Doing so leads to $P \not= \widetilde{P}$, $t_2 \not= \tilde{t}_2$ and $M \not= 0$ (as defined in  (\ref{Meq})).

\begin{figure}
\includegraphics[scale=.38]{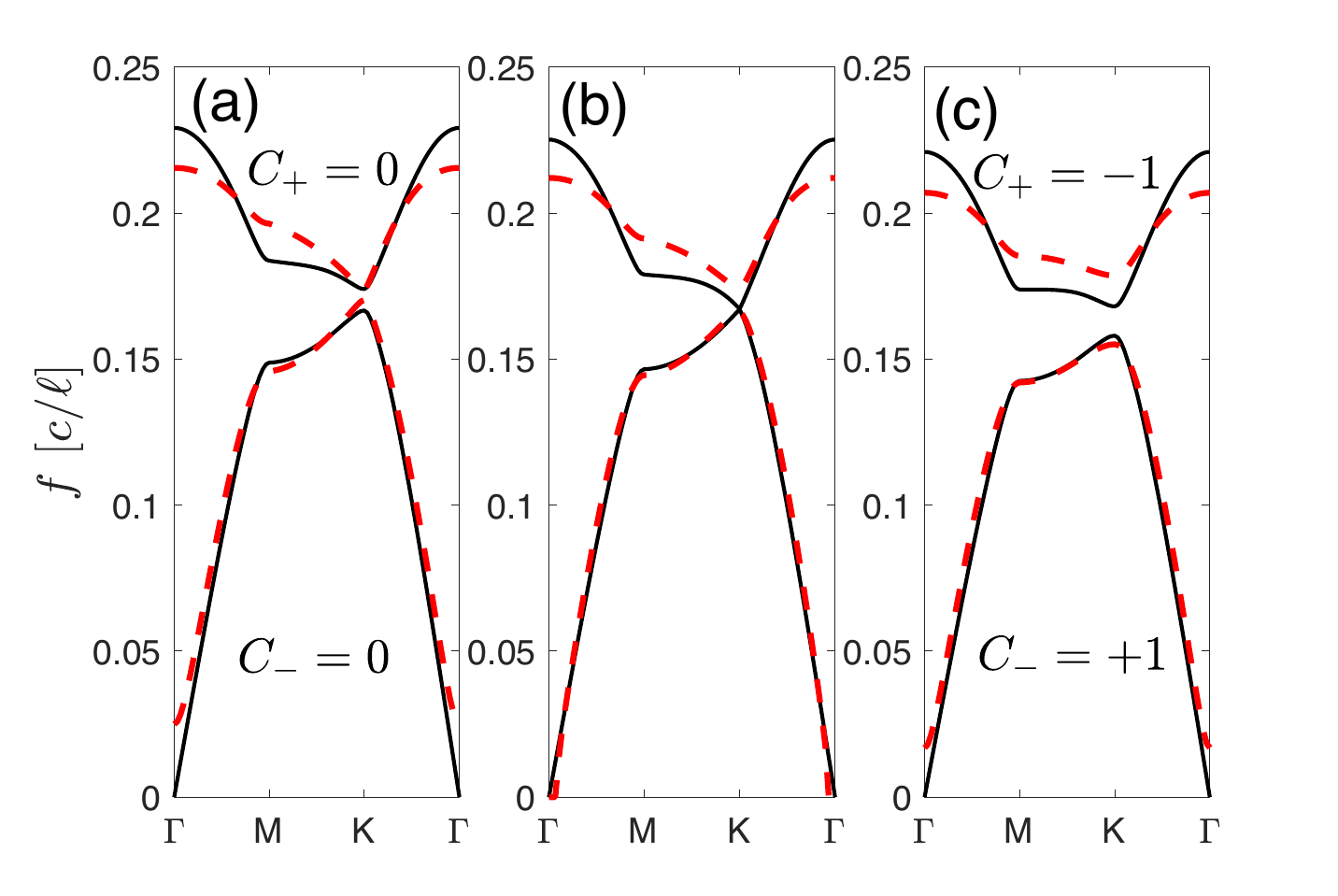}
 \caption{Topological transition of frequency Bloch bands as inversion symmetry is broken by decreasing the radius $R_a$ while keeping $R_b= 0.3$. 
 Numerical (solid curves) and tight-binding (dashed curves) approximations are shown for (a) $R_a = 0.25\ell, (b) R_a = 0.27\ell, \text{~and~} (c) R_a = 0.3\ell$. \label{Break_Inv_Sym_GMK_path}}
\end{figure}

\begin{figure}
\centering
\includegraphics[scale=.38]{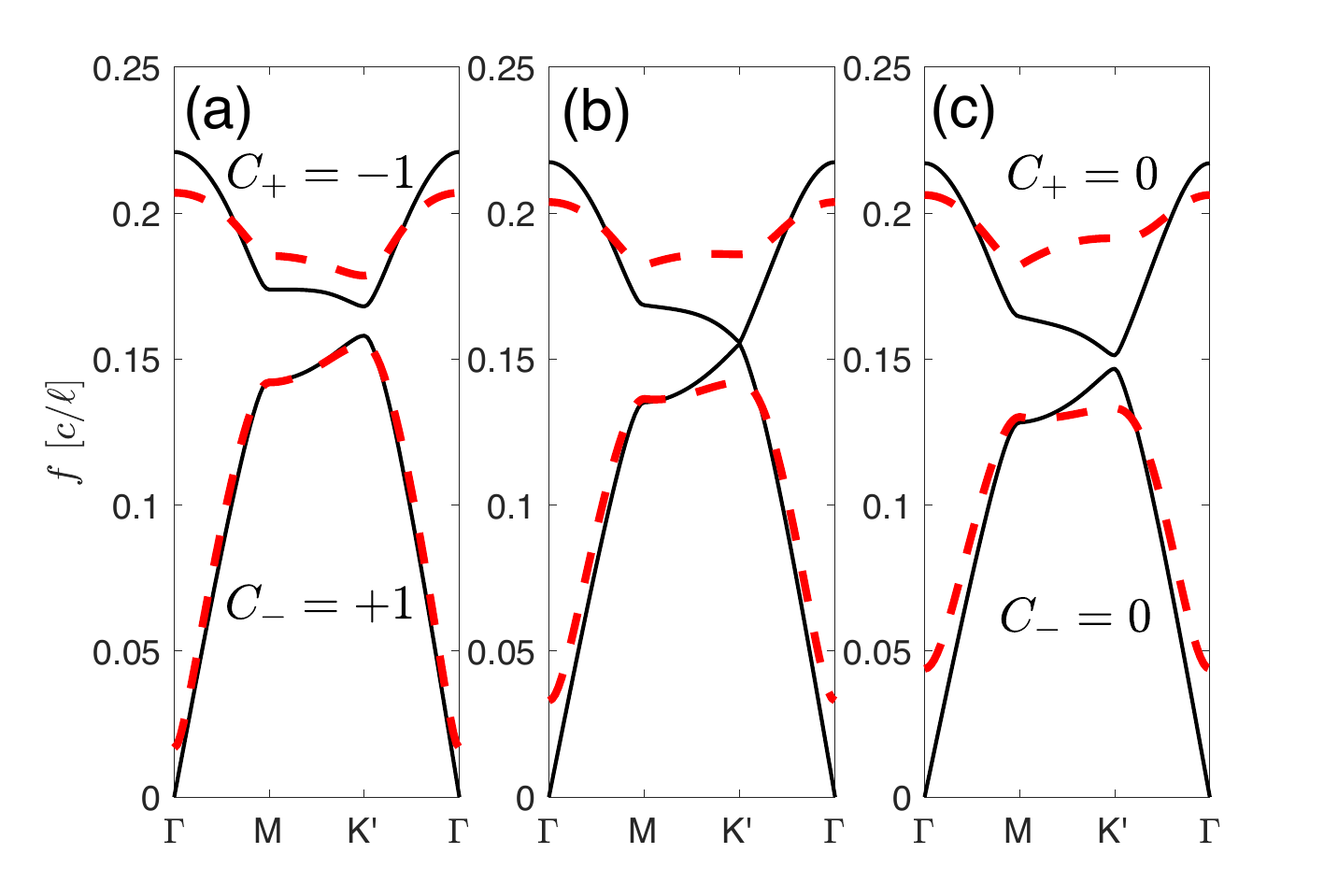}
 \caption{Spectral bands and topological transition when inversion symmetry is broken by increasing the radius $R_a$ while fixing $R_b=0.3\ell$. Numerical (solid curves) and tight-binding (dashed curves) approximations are shown for (a) $R_a = 0.3 \ell,$ (b) $R_a = 0.35 \ell,$ and (c) $R_a = 0.4 \ell.$ 
 \label{Break_Inv_Sym_GMKp_path}}
\end{figure}

Spectral band diagrams resulting from such a change are shown in Figs.~\ref{Break_Inv_Sym_GMK_path} and \ref{Break_Inv_Sym_GMKp_path}. As the radii differential changes, the system undergoes a topological transition that is captured by the model. We observe that when $R_a$ is sufficiently smaller (or larger) than $R_b$,  the system is a trivial Chern insulator. Only when $R_a \approx R_b$ do we observe a nontrivial Chern insulator state.

Specifically, for fixed radius $R_b = 0.3 \ell$, the (numerical) transition points between a trivial and nontrivial  Chern insulator occurs at approximately $R_a = 0.27 \ell$ (at $k=K$; see Fig.~\ref{Break_Inv_Sym_GMK_path}(b)) and $R_a = 0.35 \ell$ (at $k=K'$; see \ref{Break_Inv_Sym_GMKp_path}(b)). 
The discrete model (\ref{Haldane1})-(\ref{Haldane2}) is also found to exhibit these topological transitions when the inversion symmetry is broken, i.e. $M \not= 0 $ in (\ref{Meq}). 
The different regions of  topology  were  analytically studied in Sec.~\ref{analytical_topology_calc}; this information is summarized in Fig.~\ref{gap_touch_region}. 

 We note that for values of $\tau$ smaller than 1, like Fig.~\ref{Break_Inv_Sym_GMK_path}, typically the difference $|R_a- R_b|$ for $R_a < R_b$ needs to be smaller
to see a topological transition (numerical bands touch for $R_a = 0.27\ell , R_b = 0.3 \ell$, so $|R_a - R_b| = 0.03 \ell$).
In contrast, when $R_a > R_b$ and $\tau$ is larger than 1, like Fig.~\ref{Break_Inv_Sym_GMKp_path}, a larger difference $|R_a - R_b|$ is needed for a topological transition (numerical bands touch for $R_a = 0.35\ell , R_b = 0.3 \ell$, so $|R_a - R_b| = 0.05 \ell$). In examining the locations of the parameters (see Table~\ref{TB_parameters} in Appendix \ref{tight_binding_parameters}) relative to the topological regions shown in Fig.~\ref{gap_touch_region}, it appears this source of the asymmetry is the noticeable change in the value $\phi$ as $R_a$ increases.
 This differs from the behavior when $R_a$ decreases, where $\phi$ does not change substantially. This asymmetry in the transition points also occurs  if instead $R_a$ is fixed and $R_b$ is adjusted. The main difference is that the spectral touching points switch from what was observed above: $K \leftrightarrow K'$.

\section{Topologically Protected Edge Modes} 

\begin{figure}
\includegraphics[scale=.42]{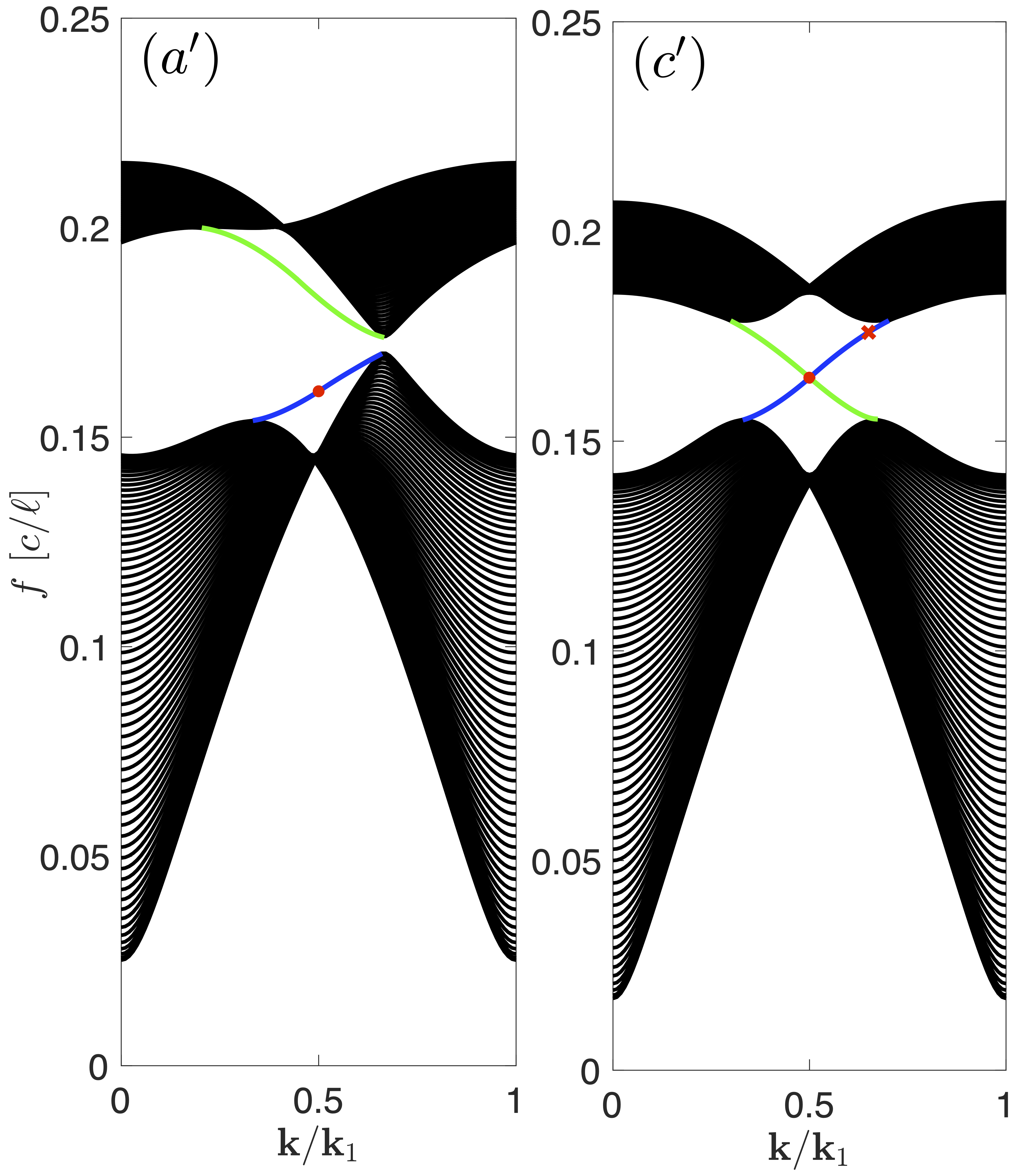}
 \caption{Spectral edge bands parallel to the ${\bf v}_1$ zig-zag edge. The model parameters for the (a$'$) topologically trivial
 and (c$'$) topologically nontrivial band diagrams are the same as those in Figs.~\ref{Break_Inv_Sym_GMK_path}(a) and \ref{Break_Inv_Sym_GMK_path}(c), respectively. Green (blue) curves correspond to edge modes localized along the top (bottom) boundary, decaying in the $+{\bf v}_2$ ($- {\bf v}_2$) direction. \label{Edge_bands_top_vary}}
\end{figure}

The edge problem is now considered. An edge is placed along the zig-zag edge parallel to the ${\bf v}_1$ lattice vector. 
Outside a semi-infinite strip, the electric field is assumed to decay exponentially fast. We find edge states that decay exponentially  in the ${\bf v}_2$ direction. 
Two topologically distinct edge band diagrams are shown in Fig.~\ref{Edge_bands_top_vary}. Edges modes along the direction ${\bf v}_1$ are found
by taking
\begin{equation*} 
a_{mn}(t) = a_n({\bf k}) e^{i \left[ m {\bf k} \cdot  {\bf v}_1   + \omega t \right]}  , ~ b_{mn}(t) = b_n({\bf k}) e^{i \left[ m {\bf k} \cdot  {\bf v}_1  + \omega t \right]} ,
\end{equation*}
which reduces the governing system (4)-(5) to
\begin{align} 
& \omega^2 a_n = P a_n + t_1 \left[ (1 + e^{ - i {\bf k} \cdot {\bf v}_1}) b_n + b_{n-1} \right]  \\ \nonumber
 +& t_2 e^{i \phi} \left[ a_{n+1} + e^{ - i {\bf k} \cdot {\bf v}_1} a_n + e^{ i {\bf k} \cdot {\bf v}_1} a_{n-1} \right]  \\  \nonumber + &  t_2 e^{-i \phi} \left[ a_{n-1} + e^{  i {\bf k} \cdot {\bf v}_1} a_n + e^{ - i {\bf k} \cdot {\bf v}_1} a_{n+1} \right] = 0 , \\ \nonumber \\ 
 & \omega^2 b_n = \widetilde{P} b_n + t_1 \left[ (1 + e^{  i {\bf k} \cdot {\bf v}_1}) a_n + a_{n+1} \right]  \\ \nonumber
 +& \tilde{t}_2 e^{-i \phi} \left[ b_{n+1} + e^{ - i {\bf k} \cdot {\bf v}_1} b_n + e^{ i {\bf k} \cdot {\bf v}_1} b_{n-1} \right]  \\ \nonumber +& \tilde{t}_2 e^{i \phi} \left[ b_{n-1} + e^{  i {\bf k} \cdot {\bf v}_1} b_n + e^{ - i {\bf k} \cdot {\bf v}_1} b_{n+1} \right] = 0 .
\end{align}
Note that ${\bf k} \cdot {\bf v}_1 = ( r {\bf k}_1 + s {\bf k}_2) \cdot {\bf v}_1 =  2 \pi r$ for $r,s \in \mathbb{R}$ due to the relationship ${\bf k}_i \cdot {\bf v}_j = 2 \pi \delta_{ij}$. As a result, the coefficients cover one period over
$0 \le r \le 1$.
This system is solved numerically by implementing zero Dirichlet boundary conditions
$$ a_n, b_n = 0 , ~~~~ n < 0, n > N$$
where $N$ is taken to be large.
We took $N = 64$ to generate Fig.~\ref{Edge_bands_top_vary}. The band gap eigenfunctions  are found to be exponentially localized and decay rapidly away from the boundary wall, in the ${\bf v}_2$ direction.

The band configuration in Fig.~\ref{Edge_bands_top_vary}(a$'$) corresponds to bulk eigenmodes with  zero Chern number due strong  inversion symmetry breaking. As a result, there are no edge modes spanning the entire frequency gap. On the other hand, the system with corresponding nonzero Chern numbers in Fig.~\ref{Edge_bands_top_vary}(c$'$) exhibits a nontrivial 
band structure inside the gap. These topologically protected chiral states propagate unidirectionally.

Finally, we consider time evolutions of these topologically distinct states. 
To do so, envelope approximations are evolved by taking the quasi-monochromatic initial data
\begin{align*}
& a_{mn}(0) = {\rm sech}(\nu m) e^{i m \overline{\bf k} \cdot {\bf v}_1} a_n(\overline{\bf k})  \\
 & b_{mn}(0) = {\rm sech}(\nu m) e^{i m \overline{\bf k} \cdot {\bf v}_1} b_n(\overline{\bf k}) ,
\end{align*}
where $a_n,b_n$ are numerically computed edge states   indicated by the red dots at $ \overline{\bf k} = 0.5 {\bf k}_1$ in Fig.~\ref{Edge_bands_top_vary} and $\nu$ is a relatively small parameter; here we took $\nu = 0.1$. 
Edge eigenmodes corresponding to localized along the bottom edge of MO honeycomb lattice are taken. 
 The edge envelopes are then propagated into a defect barrier missing two lattice cells in the $-{\bf v}_2$ direction, in which the electric field is negligibly small. 
 
\begin{figure}
  \centering 
  \includegraphics[scale=.41]{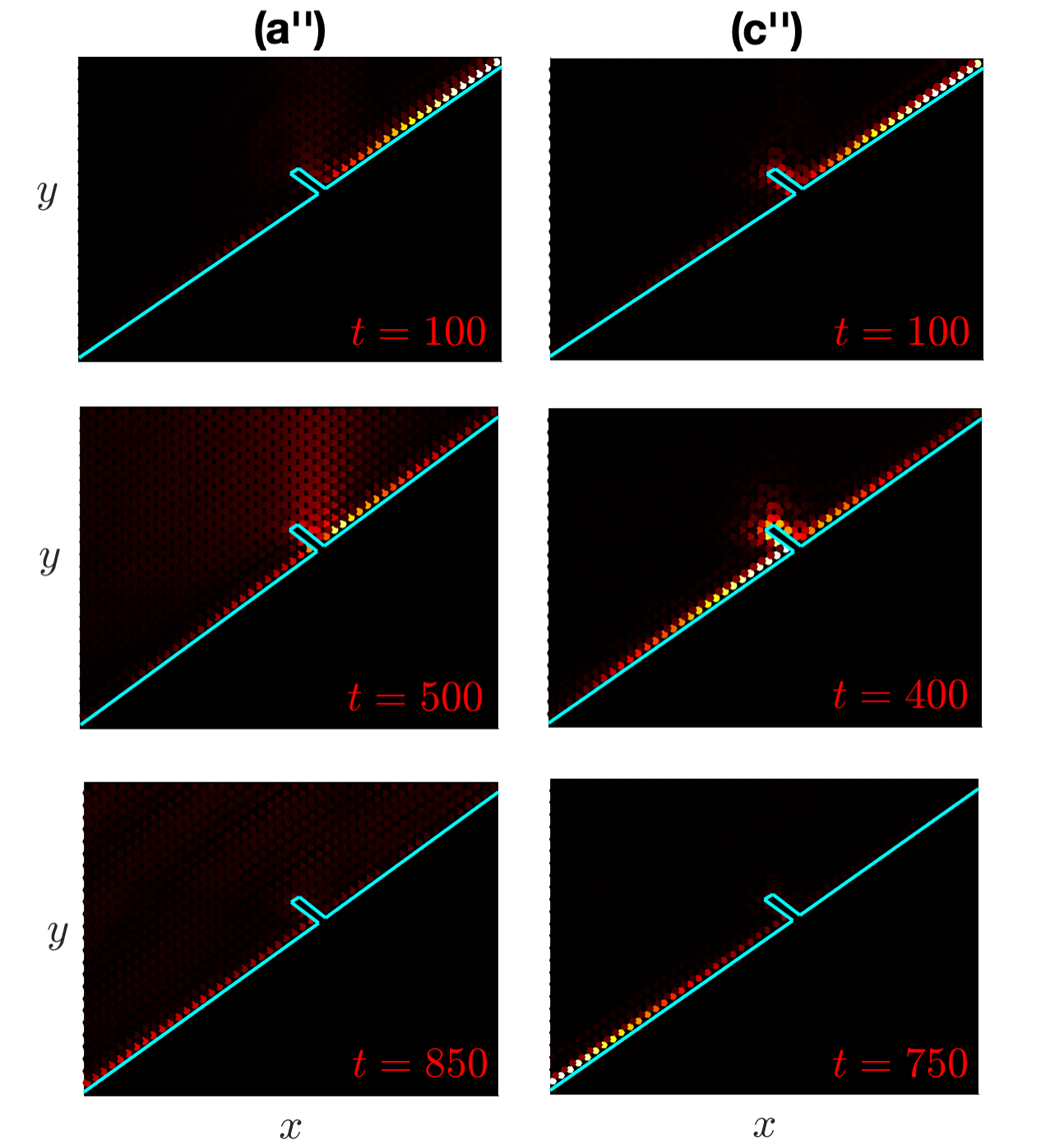}
  \caption{Envelope evolutions along and through a defect barrier. Shown are both magnitudes $|a_{mn}(t)|$ and $|b_{mn}(t)|$ at times in units of $\ell/c$. Shown are (a$''$) trivial and (c$''$) nontrivial Chern insulators with band diagrams given in Figs.~\ref{Edge_bands_top_vary}(a$'$) and \ref{Edge_bands_top_vary}(c$'$), respectively. Note: brightness is relative to the $t=100$ magnitude.  The boundary is illustrated in teal color.
  \label{evolution_compare_adjust_reduce}} 
\end{figure}

Using the initial condition above, the evolutions obtained by solving (\ref{Haldane1})-(\ref{Haldane2}) are highlighted in Fig.~\ref{evolution_compare_adjust_reduce}.
Edge states with corresponding nontrivial Chern invariants (see Figs.~\ref{Edge_bands_top_vary}(c$'$) and \ref{evolution_compare_adjust_reduce}(c$''$)) propagate chiraly around the defect barrier. There is virtually no loss in amplitude. On the other hand, edge modes associated with zero Chern number (see Figs.~\ref{Edge_bands_top_vary}(a$'$) and \ref{evolution_compare_adjust_reduce}(a$''$)) experience significant losses and scattering upon collision with the barrier. A portion of the original envelope propagates around the barrier, but there is a nearly 67\% 
amplitude loss due to scattering into the bulk.

As a final note, we observe small decay in the maximal amplitude of the these topologically protected modes, a roughly $10 \%$ decline over $1500$ time units. This is expected due to dispersion. It is well-known that  self-focusing nonlinearity can balance these dispersive effects and form solitons \cite{Ablowitz_book}. This motivates this next section which investigates a nonlinear Haldane model and edge solitons.

\section{A Nonlinear Haldane Model} 

In this section, we consider the effects of nonlinearity in our Haldane model. Similar versions of this model were mentioned in the Introduction section. The physical motivation here is that of a (relatively) high power electric field with non-negligible third-order polarization effects. The result is an onsite Kerr-type term that is proportional to the field intensity.

We consider the following nonlinear Haldane model 
\begin{align}
\label{NL_Haldane1}
\frac{d^2 a_{mn}}{dt^2} & + P a_{mn} + t_1 (\delta_- b_{mn} )  + \sigma |a_{mn}|^2 a_{mn} \\ 
\nonumber & + t_2  e^{i \phi} ( \Delta_1  a_{mn} ) +  t_2 e^{-i \phi} ( \Delta_2  a_{mn} ) = 0 \\
\label{NL_Haldane2}
\frac{d^2 b_{mn}}{dt^2} & + \widetilde{P} b_{mn} + t_1 (\delta_+ a_{mn} ) + \sigma |b_{mn}|^2 b_{mn} \\
\nonumber & + \tilde{t}_2 e^{-i {\phi}} ( \Delta_1 b_{mn} )+ \tilde{t}_2 e^{i {\phi}} ( \Delta_2  b_{mn}) = 0 
\end{align}
where the linear interaction coefficients, $\delta_{\pm},\Delta_{j}, j=1,2$, 
are defined  below (\ref{Haldane1})-(\ref{Haldane2}). Motivated by previous studies, we take an on-site focusing, Kerr-type nonlinearity, i.e. $\sigma > 0$. For the simulations below, we take $\sigma = 0.1$. 

The initial conditions used to generate solitons below  are of the form
\begin{align*}
& a_{mn}(0) =  A~ {\rm sech}(\nu m) e^{i m \overline{\bf k} \cdot {\bf v}_1} a_n(\overline{\bf k})  \\
 & b_{mn}(0) = A ~{\rm sech}(\nu m) e^{i m \overline{\bf k} \cdot {\bf v}_1} b_n(\overline{\bf k}) ,
\end{align*}
with $\overline{\bf k} = 0.65 {\bf k}_1$.
We choose a linear edge mode whose corresponding dispersion (second-derivative) is nonzero, i.e. $\omega'' \left(\overline{\bf k} \right) \approx -0.287  < 0 $ (see red `x' marker in Fig.~\ref{Edge_bands_top_vary}). Note that these derivatives are defined in the directional derivative sense
\begin{equation*}
\omega'({\bf k}) = 
\nabla \omega |_{\bf k_1}= \lim_{h \rightarrow 0} \frac{\omega ( {\bf k} + h {\bf k}_1) - \omega ( {\bf k})}{h} .
\end{equation*}
For reference, the group velocity is $\omega' \left(\overline{\bf k} \right) \approx -0.0584$ in the ${\bf v}_1$ direction. 
 Unfortunately, the third order dispersion is relatively large, $\omega''' \left(\overline{\bf k} \right) \approx -1.401$ which will  impact the formation of solitons. For this relatively weak dispersion, we seek a comparable weak nonlinearity to balance it, i.e. $A = 0.3$. A corresponding slowly-varying profile ($\nu = 0.15$) is chosen to ensure as pure of a single edge mode as possible is excited.
 
\begin{figure}
  \centering 
  \includegraphics[scale=.38]{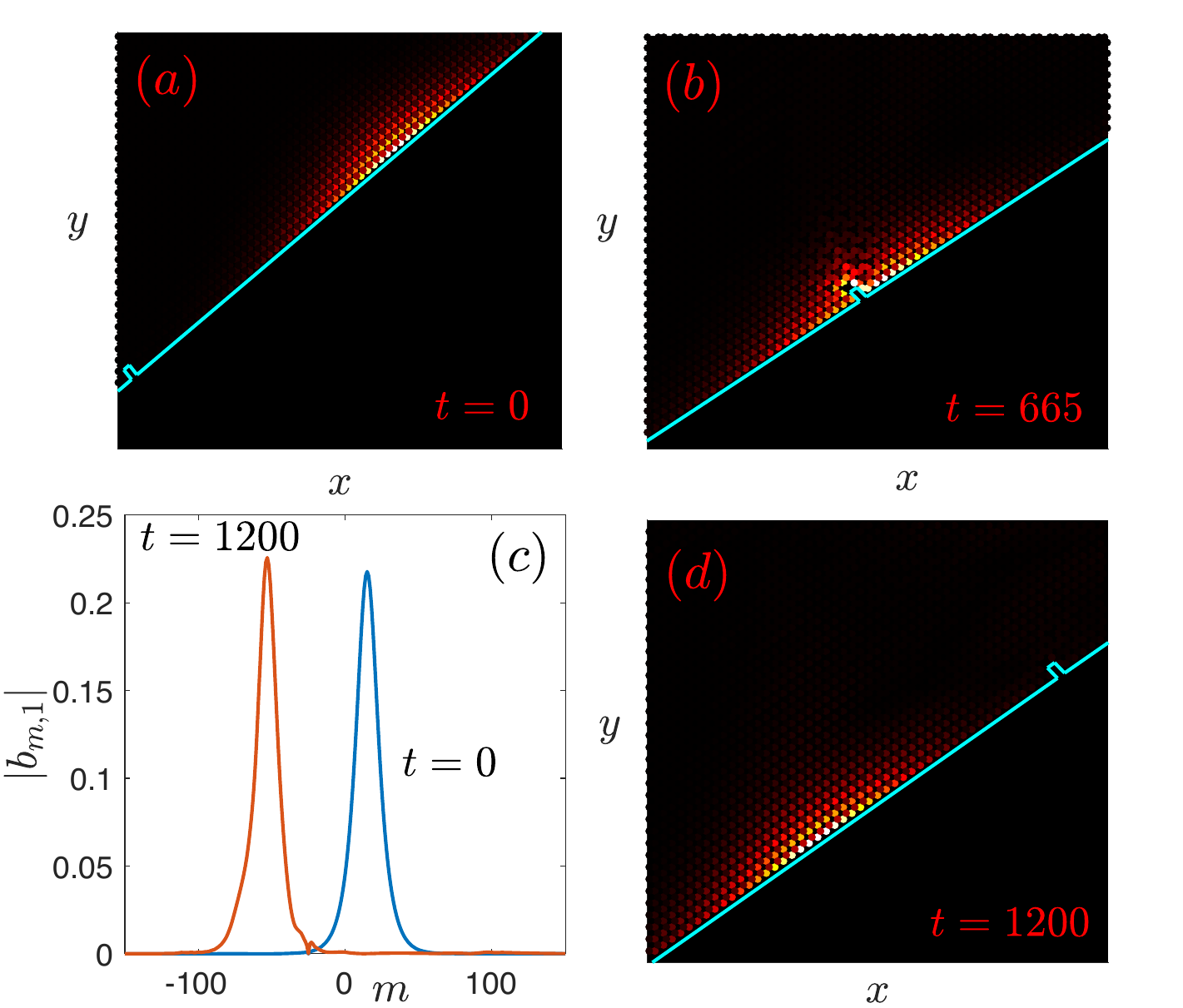}
  \caption{Nonlinear edge soliton evolutions through a defect. 
  Shown in panels (a,b,d) are both magnitudes $|a_{mn}(t)|$ and $|b_{mn}(t)|$ at times in units of $\ell/c$. Note: brightness is relative to the $t=0$ magnitude.  The boundary is illustrated in teal color; the defect location is fixed. In panel (c) the edge profile at the initial and final times is shown.
  \label{NL_defect_summary}} 
\end{figure}

Using the parameters described above, a typical evolution thorough a one 
lattice cell defect in the $- {\bf v}_2$ direction is highlighted in Fig.~\ref{NL_defect_summary}.
The resulting nonlinear mode propagates over relatively long time scales ($0 \le t \le 1200$) with a nearly constant solitary form. We observe a small $3.7 \%$ relative change in maximum magnitude between the initial and final states. 
Hence, we refer to this as an edge soliton. We note that, eventually, on longer time scales higher-order dispersion terms will become non-negligible and the mode will degrade. 

Now, some further remarks 
about these topologically protected  edge solitons. Choosing the appropriate 
sign of dispersion is imperative for achieving a self-focusing effect and solitons. When $\omega''({\bf k}) > 0$, we observe a gradual self-defocusing dissipation of the envelope. The ideal scenario for soliton formation in this MO lattice is a when $\omega''({\bf k}) <0$ and $\omega'''({\bf k}) \approx 0$. Choosing a mode centered at the zero-dispersion (inflection) point, i.e. $\omega''({\bf k}) = 0$, results in substantial dispersive break-up of initially localized solitary waves. 

In the nonlinear case, when we send a slowly-modulated mode corresponding to topologically trivial (null Chern number) into a defect, we observe significant radiation into the bulk, similar to that observed in Fig.~\ref{evolution_compare_adjust_reduce}(c'') for 
the linear evolution. In this weakly nonlinear regime, it is important to modulate a topologically nontrivial linear mode to obtain robust, unidirectional propagation. For a fully robust edge soliton, this nontrivial topology must be paired with a balanced soliton envelope.

\section{Conclusion} 

 A perturbed Wannier approach for obtaining  tight-binding approximations containing nearest and next-nearest neighbors   of a magneto-optical honeycomb lattice system is 
studied. 
Remarkably, this method leads to 
the celebrated system studied by Haldane in 1988 \ref{Haldane1}. This model agrees with experiments \cite{Poo2011} and  indicates topological transitions can occur when inversion symmetry 
and  time-reversal symmetry are broken. This data-driven Wannier approach has been previously employed in rectangular lattice geometries \cite{Ablowitz2020} and can be applicable for discovering and extrapolating discrete reductions in other Chern insulator systems in cases where a direct Wannier approach is ineffective. \\

{\bf Acknowledgements}
This project was partially supported by AFOSR under grants No. FA9550-19-1-0084 and No. FA9550-23-1-0105.

\appendix

\section{Numerical Computation of Spectral Bands and Wannier Functions} 
\label{numerical_compute_bands}

The numerical computation of the Bloch modes and maximally localized Wannier functions (MLWFs) is reviewed below. A more comprehensive discussion  can be found in \cite{Ablowitz2020}.
To simplify the necessary calculations, first, a linear transformation is introduced to map the parallelogram unit cell to a square. The change of variables
\begin{equation*}
u = \frac{1}{3} x - \frac{1}{\sqrt{3}} y  , ~~~~~~ v = \frac{1}{3} x + \frac{1}{\sqrt{3}} y 
\end{equation*}
transforms the parallelogram formed by the lattice vectors ${\bf v}_1, {\bf v}_2$ into a square with side length 1. As a result, the master equation (\ref{TM_eqn}) transforms to
\begin{align}
\label{transformed_TMeqn}
 \nonumber - \frac{4}{9} \left( \frac{\partial^2 E}{ \partial u^2} - \frac{\partial^2 E}{ \partial u \partial v} + \frac{\partial^2 E}{ \partial v^2} \right) &+ g(u,v) \frac{\partial E}{ \partial u} + h(u,v) \frac{\partial E}{ \partial v} \\  & = \omega^2 \varepsilon(u,v) \tilde{\mu}(u,v) E 
\end{align}
where
\begin{align*}
& g(u,v) = \frac{4}{9\tilde{\mu}} \frac{\partial \tilde{\mu}}{\partial u} - \frac{2}{9\tilde{\mu}} \frac{\partial \tilde{\mu}}{\partial v} + \frac{2 i \tilde{\mu}}{3 \sqrt{3}} \frac{\partial \eta}{ \partial v} \\ \nonumber
& h(u,v) =-  \frac{4}{9\tilde{\mu}} \frac{\partial \tilde{\mu}}{\partial u} + \frac{2}{9\tilde{\mu}} \frac{\partial \tilde{\mu}}{\partial v} - \frac{2 i \tilde{\mu}}{3 \sqrt{3}} \frac{\partial \eta}{ \partial v} .
\end{align*}
From here, a formulation similar to that used in \cite{Ablowitz2020} can be applied. All transformed coefficients now have the periodicity: $f(u + m  , v + n ) = f(u,v)$ for $m,n \in \mathbb{Z}$. That is, the functions are periodic with respect to the transformed lattice vectors ${\bf e}_1 = (1 , 0)^T$ and ${\bf e}_2 = (0,1)^T$.
Hence, transformed master equation (\ref{transformed_TMeqn}) is solved by looking for Bloch wave solutions with the form $E({\bf w}, {\bm \kappa}) = e^{i {\bm \kappa} \cdot {\bf w}} u({\bf w}; {\bm \kappa})$, where $u({\bf w} + m {\bf e}_1 + n {\bf e}_2; {\bm \kappa}) = u({\bf w}; {\bm \kappa})$ for ${\bf w} = (u,v)^T, {\bm \kappa} = (k_u, k_v)^T$. Note: this ${\bm \kappa}$ is unrelated to $\kappa$ of the permeability tensor. 

The numerical spectral bands shown throughout this paper are computed by solving (\ref{transformed_TMeqn}) for the eigenfunction/eigenvalue pair $(u,\omega)$ as functions of the transformed quasimomentum. Subsequently, the quasimomentum is transformed back to the original $k_x,k_y$ variables via
\begin{equation*}
k_x = \frac{k_u + k_v}{3} , ~~~~~~~ k_y = \frac{k_v - k_u}{\sqrt{3}} .
\end{equation*}
The continuous Chern numbers are defined by
\begin{equation*}
\label{chern_define}
C_p =   \frac{1}{2 \pi i} \iint_{\rm BZ} (\nabla_{\bf k} \times {\bf A}_p) \cdot \widehat{{\bf z}} ~ d{\bf k} ~ ,
\end{equation*}
where ${\bf A}_p({\bf k}) = \langle u_p({\bf r}, {\bf k}) | \partial_{k_x} u_p({\bf r}, {\bf k})   \rangle_{{\rm UC},\epsilon \tilde{\mu} }~  \widehat{\bf x} + \langle u_p({\bf r}, {\bf k}) | \partial_{k_y} u_p({\bf r}, {\bf k})   \rangle_{{\rm UC},\epsilon \tilde{\mu}}   ~ \widehat{\bf y} $
are numerically computed using the algorithm \cite{fukui} with respect to the weighted inner product
\begin{equation*}
\langle f, g \rangle_{{\rm UC},\epsilon \tilde{\mu} } =  \iint_{\rm UC} f({\bf r})^* g({\bf r}) \varepsilon({\bf r}) \tilde{\mu}({\bf r}) d {\bf r} ,
\end{equation*}
where ${\rm UC}$ denotes the unit cell.

The transformed Bloch wave is periodic with respect to the transformed reciprocal lattice vectors, ${\bm \kappa}_1 = 2 \pi (1 , 0)^T, {\bm \kappa}_2 = 2 \pi  (0 ,1)^T$. Notice: ${\bf e}_i \cdot {\bm \kappa}_j = 2 \pi \delta_{ij}$. As such, it can be expressed as the Fourier series
\begin{equation*}
E({\bf w}; {\bm \kappa}) = \sum_{p} \sum_{m,n}  W_{mn}^p({\bf w}) e^{i {\bm \kappa} \cdot (m {\bf e}_1 + n {\bf e}_2) } ,
\end{equation*}
where $W_{mn}^p({\bf w})$ is a transformed Wannier function corresponding to the $p^{\rm th}$ band, and centered at the $(m,n)$ unit cell.
For the problem studied in this paper, we only consider the lowest two bands $p = 1,2$ and truncate the remaining terms. 

Next, the MLWF algorithm \cite{Marzari1997} is applied to find localized Wannier functions for the $g = h = 0$ (corresponding to $\mathcal{M} = {\bf 0}$) problem in equation (\ref{transformed_TMeqn}). This is done by finding a unitary transformation that minimizes the functional describing the variance of the Wannier function, given by equation (\ref{variance_define}) below. Let $E_1({\bf w}, {\bm \kappa}) = e^{i {\bm \kappa} \cdot {\bf w}} u_1({\bf w}; {\bm \kappa})$  and $E_2({\bf w}, {\bm \kappa}) = e^{i {\bm \kappa} \cdot {\bf w}} u_2({\bf w}; {\bm \kappa})$ correspond to first and second spectral bands, respectively.
A spectral unitary transformation  of the Bloch functions is taken at fixed values of ${\bf w}$ 
\begin{equation*}
\begin{pmatrix}
\widetilde{u}^c({\bf w}; {\bm \kappa}) \\
\widetilde{u}^d({\bf w}; {\bm \kappa})
\end{pmatrix}
=
\mathbb{U}({\bm \kappa})
\begin{pmatrix}
{u}_1({\bf w}; {\bm \kappa}) \\
{u}_2({\bf w}; {\bm \kappa})
\end{pmatrix}
\end{equation*}
where $\mathbb{U}({\bm \kappa})$ is a $2\times 2$ matrix. Only after computing these Wannier functions do we realize where they are physically located. Upon inspection, we replace the labels $c,d$ with the labels $a,b$, where $a$ modes are centered at the $a$-sites and $b$ modes centered at the $b$-sites (see Fig.~\ref{Wannier_Fcn_Plot}).

Upon obtaining these functions, the Bloch modes $\widetilde{E}^a({\bf w}; {\bm \kappa})  = e^{i {\bm \kappa} \cdot {\bf w}} \widetilde{u}^a({\bf w}; {\bm \kappa})$ and $ \widetilde{E}^b({\bf w}; {\bm \kappa}) = e^{i {\bm \kappa} \cdot {\bf w}} \widetilde{u}^b({\bf w}; {\bm \kappa}) $ are computed and then  used to construct the Wannier functions
\begin{align*}
&\widetilde{W}^a_{mn}({\bf w}) = \frac{1}{4 \pi^2} \iint_{\rm BZ} e^{ - i {\bm \kappa} \cdot (m {\bf e}_1 + n {\bf e}_2 )} \widetilde{E}^a({\bf w}; {\bm \kappa}) d{\bm \kappa} \\
&\widetilde{W}^b_{mn}({\bf w})  = \frac{1}{4 \pi^2} \iint_{\rm BZ} e^{ - i {\bm \kappa} \cdot (m {\bf e}_1 + n {\bf e}_2 )} \widetilde{E}^b({\bf w}; {\bm \kappa}) d{\bm \kappa}
\end{align*}
shown in  Fig.~\ref{Wannier_Fcn_Plot}. 

\section{Tight-binding Parameters}
\label{tight_binding_parameters}

The parameters used to produce the tight-binding approximations are given in Table~\ref{TB_parameters}. All cases correspond to the magnetization by an external field, or $\mathcal{M}({\bf r}) \not= {\bf 0}$, except  (i) which is the unmagnetized case. Also included are the corresponding rod radii. 
The Chern number corresponding to the upper spectral surface of the tight-binding model is included. 
The value $C_+ = -1$ 
corresponds to phase points located inside the topological region of Fig.~\ref{gap_touch_region}, while $C_+ = 0$ 
lies above or below this region. The values in the table were computed analytically as well as numerically using the algorithm in \cite{fukui} on the discrete eigenvectors.
In all cases considered, the values agreed. The topological numbers for the discrete (tight-binding) model also match those for the continuum model, except possibly near the sensitive  topological transition points.
\begin{table}
\centering
\begin{tabular}{ |c|c|c|c|c|c|c|c|c|c|c|} 
 \hline
& $R_a$  & $R_b$ & $P$ & $\widetilde{P}$ &  $t_1$ & $t_2$ & $\tilde{t}_2$ & $\phi$ & $C_+$   \\ \hline
(i)  & $0.3 \ell$ & $0.3 \ell$  & 0.882 & 0.882 &  -0.262 & 0.012 &0.012 & $\pi$ &  0  \\ \hline
(ii) & $0.3 \ell$ & $0.3 \ell$ & 1.020 & 1.020  & -0.280 & 0.041 & 0.041 & 2.327 & -1  \\ \hline
(iii) & $0.27 \ell$ &$0.3 \ell$ & 1.152  & 0.991  & -0.297 & 0.049 & 0.037 & 2.391 &    \\  \hline
(iv) & $0.25  \ell$&$0.3 \ell$ & 1.263 & 0.973  & -0.300 & 0.055 & 0.030 &  2.405 & 0  \\ \hline
(v) &$0.35 \ell$ &$0.3 \ell$ & 0.876  &  1.077 & -0.265  & 0.031 & 0.048  & 2.178 &  
  \\  \hline
(vi) &$0.4 \ell$ &$0.3 \ell$ & 0.790  & 1.132   & -0.263  & 0.024  & 0.053 & 1.941 & 0   \\ 
 \hline
\end{tabular}
 \caption{\label{TB_parameters} Tight-binding parameters in (\ref{Haldane1})-(\ref{Haldane2}) for (i) non-magnetized bands and magnetized (ii-vi) systems. The rod radii used to generate these parameters are given by $R_a$ and $R_b$ for a-sites and b-sites, respectively. The value of $C_+$ 
 indicates the Chern number corresponding to the second (upper) spectral band.} 
\end{table}

\section{Breaking of Inversion Symmetry} 
\label{appendix_break_inversion}

As discussed in Sec.~\ref{broken_inversion_symmetry},  breaking  inversion symmetry can induce a topological transition from a nontrivial 
to trivial Chern insulator. This symmetry breaking can be implemented by  choosing different radii at a-sites and b-sites, that is $R_a \not= R_b$ in Fig.~\ref{HC_lattice}.  The spectral bands induced by this change are shown in Figs.~\ref{Break_Inv_Sym_GMK_path} and \ref{Break_Inv_Sym_GMKp_path}. In particular, decreasing $R_a$  relative to $R_b$ induces a topological transition  and a touching point at the $K$ Dirac point. If, on the other hand, one considers $R_a$ larger relative to $R_b$, a similar topological transition occurs, but instead the gap closes at the opposite Dirac point, $K' = - K$.
A  depiction of this is shown in Fig.~\ref{Break_Inv_Sym_GMKp_path}. 

The computed parameters are summarized in Table~\ref{TB_parameters}. Examining the inversion parameter $M = (P - \widetilde{P})/2$, it is observed to be positive when $R_a < R_b$, and negative when $R_a > R_b$. 
For applications which seek to use these chiral edge modes, inversion symmetry should be nearly satisfied, that is, $R_a \approx R_b$. More precisely, the system supports topologically protected modes when the parameters $\widetilde{M}$ and $\phi$ are chosen to reside in the inner (topological) region of Fig.~\ref{gap_touch_region}.

\begin{figure*}
\centering
\includegraphics[scale=.35]{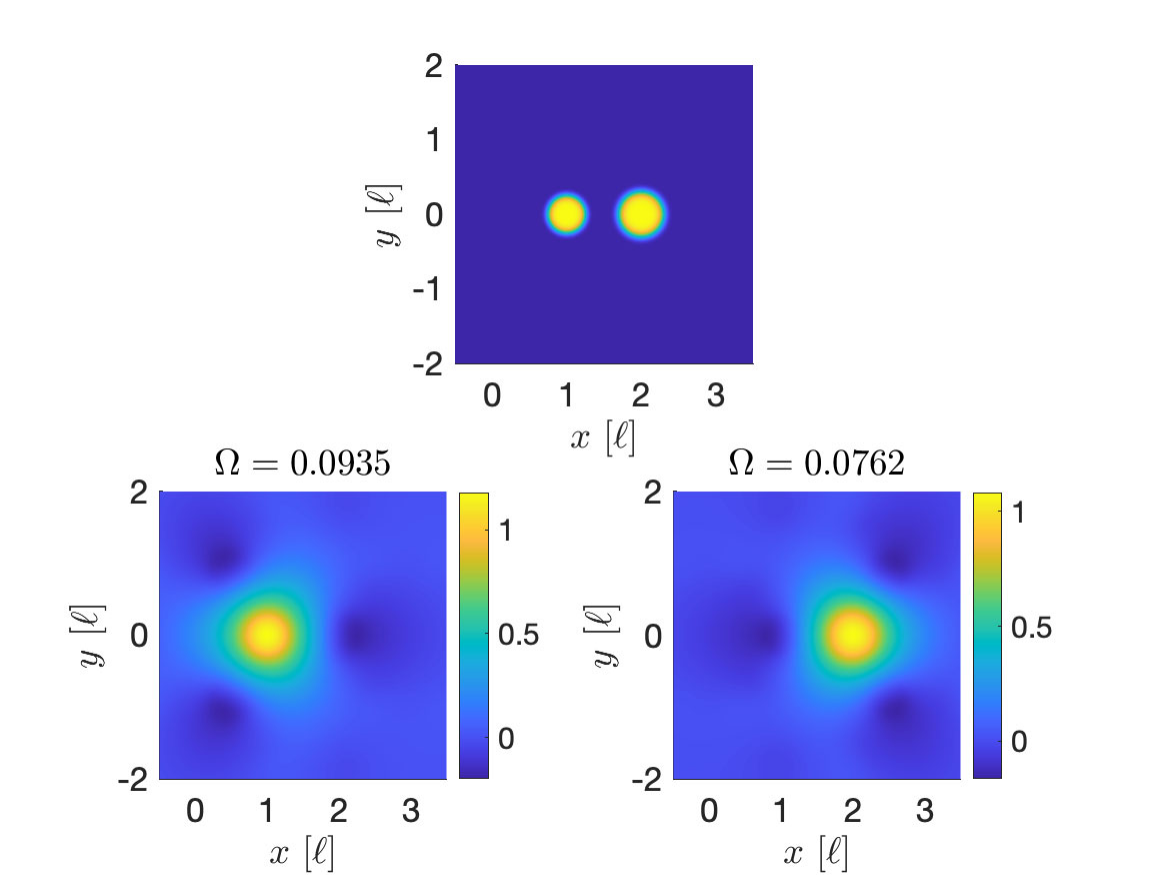}
\includegraphics[scale=.35]{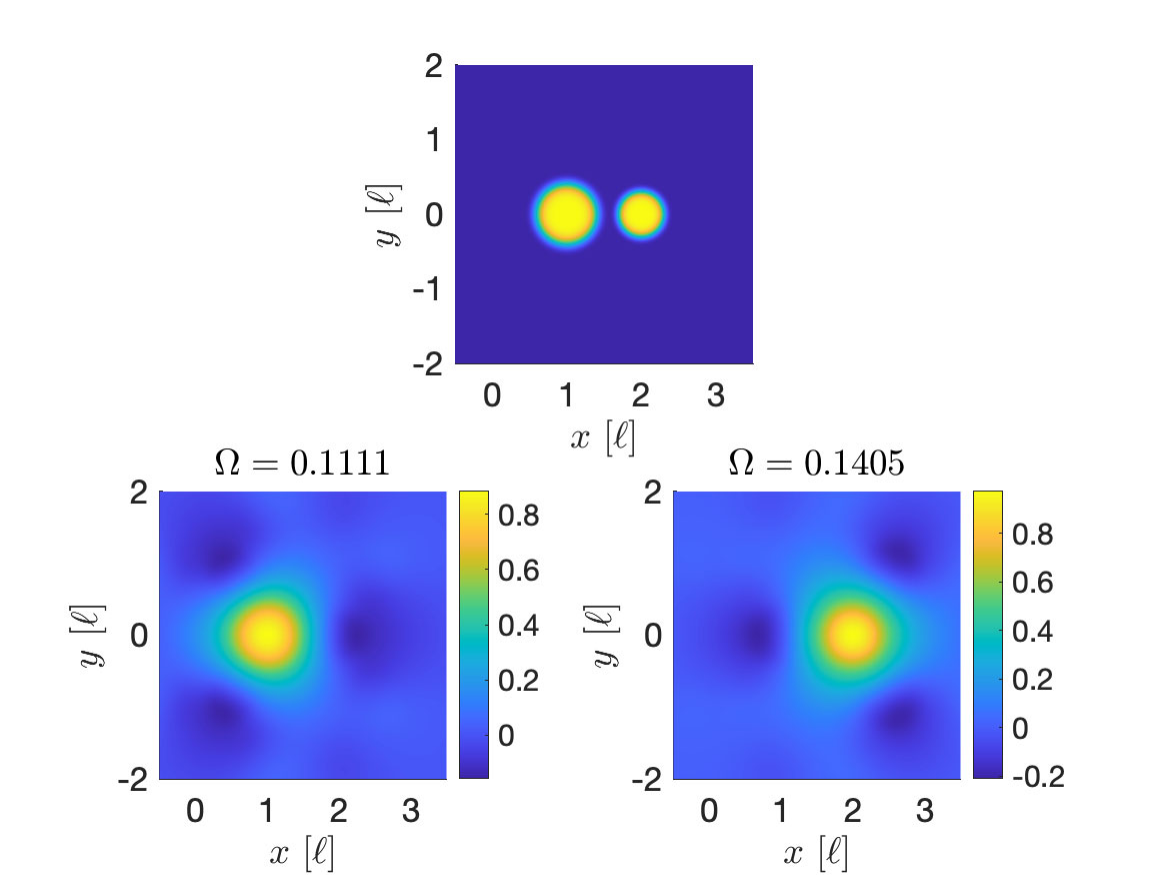}
 \caption{The top row panels highlight the rod radii for the different cases. Bottom rows show the maximally localized Wannier functions (MLWFs) corresponding to those physical parameters. (left) Radii: $R_a = 0.25 \ell, R_b = 0.3 \ell$ and (right) Radii: $R_a = 0.4 \ell, R_b = 0.3 \ell$. Also included for each Wannier function is the corresponding width, defined in (\ref{variance_define}).
 \label{Wannier_Plot}}
\end{figure*}

Some of the
Wannier functions corresponding to broken inversion symmetry are shown in Fig.~\ref{Wannier_Plot}. In each case the rod profiles and their corresponding Wannier modes are shown. These Wannier modes are constructed in a manner similar to that described in Sec.~\ref{perturbed_wannier_approach} and \cite{Ablowitz2020}, i.e. $\mathcal{M}  = {\bf 0}, \tilde{\mu} \not= \mu_0 $. Also given is the variance
\begin{equation}
 \label{variance_define}
\Omega = \langle |{\bf r}|^2 \rangle - | \langle{\bf r}\rangle |^2 ,
\end{equation}
\begin{equation*}
\langle x^n \rangle \equiv \iint_{\mathbb{R}^2} x^n |W({\bf r})|^2 \varepsilon({\bf r}) \tilde{\mu}({\bf r}) d{\bf r} 
\end{equation*}
where $W({\bf r})$ is the corresponding Wannier mode. 
The lattice sites whose rods have  {\it larger} relative radius, correlate to a {\it smaller} variances; and vice versa for smaller relative rods. This is the source of $\tilde{t}_2 \not= t_2$ and $P \not= \widetilde{P}$. These different widths indicate different decay rates and imply different tight-binding coefficients among  sites of the same type.

\end{document}